\documentclass[a4paper]{article}

\usepackage{graphicx}
\usepackage{amssymb,amsmath,amsthm}
\usepackage[multiple]{footmisc}

\newtheorem{theorem}{Theorem}
\newtheorem{remark}{Remark}
\newtheorem*{namedthm*}{\thistheoremname}
\newcommand{\thistheoremname}{} 
\newenvironment{namedthm}[1]
  {\renewcommand{\thistheoremname}{#1}\begin{namedthm*}}
  {\end{namedthm*}}

\newenvironment{AMS}{\textit{MSC2020:} }{}
\newenvironment{keywords}{\textit{key words:} }{}

\DeclareMathOperator{\DS}{DSPACE}
\DeclareMathOperator{\NS}{NSPACE}

\title{Refuting Tianrong Lin's \texttt{arXiv:2110.05942} ``Resolution of The Linear-Bounded Automata Question''}
\author{Thomas Preu\thanks{thomas[dot]preu[at]math[dot]uzh[dot]ch}}

\begin{document}
\newpage
\maketitle

\begin{abstract}
In the preprint mentioned in the title Mr.~Tianrong claims to prove $\NS[n]\neq\DS[n]$, resolving a longstanding open
problem in automata theory called the \textit{LBA question}. He claims to achieve this by showing more generally
$\NS[S(n)]\neq\DS[S(n)]$ for suitable $S(n)$. We demonstrate that his proof is incomplete, even wrong, and his strategy
cannot be repaired.
\end{abstract}



\section{Introduction}

Mr.~Tianrong uploaded his paper first on 11th Oct. 2021 to \texttt{arXiv}. Over the course of the following two
weeks we, Mr.~Tianrong and the author of this paper, had a discussion via email about open points in Mr.~Tianrong's
work. Since we could not come to agree on the open points, Mr.~Tianrong invited me to write a correct paper on my own. I
will follow his invitation and refute his claim that he resolved the LBA question.

On 21st Oct. 2021, Mr.~Tianrong uploaded an update of his paper to \texttt{arXiv} as of writing this note. We refer to
this second version \cite{LBAQ}. We mainly use the notation from there, which is standard, in particular DTM, NTM,
$\DS[S(n)]$, etc. Further updates have appeared: we will comment on that at the end of section \ref{sec:sum}.

The rest of this small note is organized in three main parts. In section \ref{sec:strat} we outline the key points of
the proof strategy for the claims in \cite{LBAQ}. In section \ref{sec:cons} we assume that the main argument
in \cite{LBAQ} was correct and based on that make some harmless modifications which would not invalidate the argument
but which lead to an obvious contradiction. By this we show that the approach of Mr.~Tianrong is fatally flawed and
is beyond repair. In section \ref{sec:cause} we make an attempt to locate the source of the error in the main argument
of \cite{LBAQ}. Finally we sum up our refute of Mr.~Tianrong claims in section \ref{sec:sum}.

\section{The strategy of Mr.~Tianrong}
\label{sec:strat}

The heart of \cite{LBAQ} is Theorem 3.~on p. 6. We repeat it here word by word:

\setcounter{theorem}{2}
\begin{theorem}
Let $S(n)\geq\log n$ be a space-constructible function. Then there exists a language $L_d$ accepted by a NTM by using
space $O(S(n))$ but by no DTM of space complexity $S(n)$. That is, $L_d\in\NS[S(n)]$ but $L_d\not\in\DS[S(n)]$.
\end{theorem}

Most of \cite{LBAQ} is inspired by \cite{AHU} which Mr.~Tianrong cites repeatedly. In particular his Theorem 3
parallels \cite[Thm. 11.1.]{AHU} on p. 408. A version with more restrictive assumptions can be found
in \cite[Thm. 12.8]{HU} on p. 297. We repeat this later statement as it seems closer to Theorem 3.

\begin{namedthm}{Theorem 12.8}
If $S_2(n)$ is a fully space-constructible function,
\[\inf\limits_{n\rightarrow\infty}\frac{S_1(n)}{S_2(n)}=0,\]
and $S_1(n)$ and $S_2(n)$ are each at least $\log_2 n$, then there is a language in $\DS(S_2(n))$ not in $\DS(S_1(n))$.
\end{namedthm}

\begin{remark}
For a TM $M$ we denote the language accepted by it by $L_M$ as is customary. Note that if a DTM $M$ uses at most $S(n)$
space, then $L_M\in\DS(S(n))$. However, even if $M$ used more space than $S(n)$, we cannot rule out $L_M\in\DS(S(n))$:
there may be another DTM $M'$ using at most $S(n)$ space and accepting the same language $L_{M'}=L_{M}$. Below we will
focus on the space requirements of a DTM $M$ and less on whether $L_M\in\DS(S(n))$.
\end{remark}

We give a two page exposition of the proof of \cite[Thm. 12.8]{HU} along the lines of the proof of 
\cite[Thm. 11.1.]{AHU}. Details and used terminology can be found in the references.

The argument in \cite[Thm. 11.1.]{AHU} is essentially based on diagonalization. First a four-tape DTM $M_0$ with
$L_{M_0}\in\DS(S_2(n))$ is constructed in five steps that on binary input $x$ of length $n=|x|$ attempts to simulate a
single-tape DTM $M=M_x$ which depends on the binary encoding $x$ (for details cf. \cite{LBAQ} or \cite{AHU}). There is a
twist to this encoding as prepending any number of $1$s to $x$ encodes the same DTM, i.e.
``$M_{x}=M_{1x}=M_{11x}=\ldots$''. Thus for any DTM $M'$ and any bound $b\in\mathbb{N}$ there is a code word $x'$ with
$M'=M_{x'}$ and $|x'|\geq b$. $M_0$ uses a binary tape alphabet with an added blank symbol\footnote{Technically
in \cite{AHU} a forth separator symbol is introduced, but this is not essential.}. The behavior of $M_0$ is defined in
these five steps:
\begin{enumerate}
\item By space-constructibility of $S_2(n)$ we can detect if $M_0$ would go beyond that bound and halt without
accepting. This enforces $L_{M_0}\in\DS(S_2(n))$.
\item If $x$ is not a code word for a DTM, $M_0$ detects this and halts without accepting. Otherwise we proceed to the
next step.
\item Then $M_0$ prepares to simulate $M_x$. For this $M_0$ determines $s$ the number of states and $t$ the length of the
tape alphabet of $M_x$. Any tape symbol for $M_x$ can be represented by $\lceil\log_2 t\rceil$ binary tape symbols of
$M_0$. We want to ensure that $L_{M_x}\in\DS(S_1(n))$. For this we limit the space of simulation to $(1+\lceil\log_2
t\rceil)S_1(n)$ tape cells\footnote{Instead we could also only use $\lceil\log_2 t\rceil S_1(n)$ as in \cite{HU}. Again
these are irrelevant details.} on $M_0$ before actually simulating $M_x$. We need to distinguish two cases:
\begin{enumerate}
\item If $(1+\lceil\log_2 t\rceil)S_1(n)\nleq S_2(n)$, then by step 1 $M_0$ halts without accepting. This can happen despite
$M_x$ not exceeding the $S_1(n)$ space bound, but this loss will be irrelevant.
\item If $(1+\lceil\log_2 t\rceil)S_1(n)\leq S_2(n)$, $M_0$ proceeds to the next step.
\end{enumerate}
\item On a separate tape $M_0$ sets up a counter of $\lceil\log_2 s\rceil+\lceil\log_2 S_1(n)\rceil+\lceil\log_2 t\rceil
S_1(n)$ many tape cells. This counter can go up to (possibly slightly more than) $sS_1(n)t^{S_1(n)}$. It will be used to
detect cycles in the configuration space of $M_x$, i.e. to detect when the simulation of $M_x$ will not
halt\footnote{This is possible because the set of DTMs respecting the space bound $S_1(n)$ do not have the full power of
the set of all DTMs -- i.e. we are not contradicting the unsolvability of the halting problem.}\footnote{In \cite{HU}
detecting non-halting behavior is outsourced to Lemma 12.1}. Each of the three
factors of $sS_1(n)t^{S_1(n)}$ is related to an essential component in the configuration space: state, position of
read-write head on tape, content of the (space-restricted) tape. Again we distinguish:
\begin{enumerate}
\item If $\lceil\log_2 s\rceil+\lceil\log_2 S_1(n)\rceil+\lceil\log_2 t\rceil S_1(n)\nleq S_2(n)$, then by step 1 $M_0$
halts without accepting. This can happen despite $M_x$ respects the $S_1(n)$ space bound and properly halts, but
this loss will be irrelevant.
\item If $\lceil\log_2 s\rceil+\lceil\log_2 S_1(n)\rceil+\lceil\log_2 t\rceil S_1(n)\leq S_2(n)$, $M_0$ proceeds to the
next step. 
\end{enumerate}
\item Here is where the heart of the diagonalization is happening. After this setup $M_0$ simulates $M_x$ on input
$x$. Its halting behavior is as follows:
\begin{enumerate}
\item If $M_0$ uses more cells than allowed by step 3, i.e. if $M_x$ uses more than $S_1(n)$ cells, then $M_0$
accepts. Note that both $L_{M_x}\in\DS(S_1(n))$ or $L_{M_x}\not\in\DS(S_1(n))$ are possible, but this distinction will
be irrelevant.
\item If the bound from step 3 is respected but the counter from step 4 overflows, i.e. if $M_x$ does not halt on
$x$, then $M_0$ accepts.  
\item If the simulation of $M_x$ on $x$ by $M_0$ respects the bounds from steps 3 and 4 and halts, then, if $M_x$ rejects,
$M_0$ will accept.
\item If the simulation of $M_x$ on $x$ by $M_0$ respects the bounds from steps 3 and 4 and halts, then, if $M_x$ accepts,
$M_0$ will reject. 
\end{enumerate}
\end{enumerate}

The four-tape DTM $M_0$ always halts and by step 1 it uses no more space than $S_2(n)$, i.e. $L_{M_0}\in\DS(S_2(n))$.
Now \cite{AHU} argues by contradiction that $L_{M_0}\not\in\DS(S_1(n))$. If otherwise, by standard lemmas (for details
cf. \cite{LBAQ} or \cite{AHU}) there would be a single-tape DTM $M_y$ using no more than $S_1(n)$ space with\footnote{We
actually use a stronger assertion, that the execution of $M_y$ runs parallel to that of $M_0$. This is however a direct
consequence of the proof of the standard lemmas.} $L_{M_y}=L_{M_0}$. So far $M_x$ was variable and with it the code word
$x$, its length $n=|x|$, the number of states $s$ and the size of the tape alphabet $t$. Now \cite{AHU} fixes $M_x=M_y$,
we will call it $\overline{M}$ from now on, and with it $s=s_{\overline{M}}$ and $t=t_{\overline{M}}$ become
constant. However the code word $z\in\{y,1y,11y,\ldots\}$ and $n=|z|$ can stay variable in a limited sense as
``$\overline{M}=M_{z}=M_{y}=M_{1y}=M_{11y}=\ldots$''.

Now everything falls into its place (again for details cf. the literature): By step 1 $L_{M_0}\in\DS(S_2(n))$. Since any
of our restricted $z\in\{y,1y,11y,\ldots\}$ encodes the fixed DTM $\overline{M}$, i.e. $z$ is a valid code word,
$\overline{M}=M_z$ executed on $z$ passes\footnote{$\overline{M}$ is a single-tape version of $M_0$ and their execution
runs in parallel. Thus we may say ``$\overline{M}$ on input $z$ passes step X'', if $M_0$ on input $z$ does not halt in
step X and proceeds further.} step 2. Then \cite{AHU} uses $\inf_{n\rightarrow\infty}\frac{S_1(n)}{S_2(n)}=0$, i.e. that
in this limited sense $S_2(n)$ grows asymptotically faster than $S_1(n)$, and they use that $s_{\overline{M}}$ and
$t_{\overline{M}}$ are now fixed constants, to pick a $w\in\{y,1y,11y,\ldots\}$ of sufficiently large size $n_w=|w|$,
s.t. $\overline{M}$ on input $w$ passes steps 3 and 4. By assumption from the last paragraph $\overline{M}=M_w=M_y$ uses
at most $S_1(n)$ space, in particular when executed on $w$, thus 5(a) is not relevant. As $\overline{M}$ is a
single-tape version of $M_0$ which always halts we also can ignore 5(b). Since only 5(c) and 5(d) are left, $M_w$
accepts $w$ iff $M_0$ rejects $w$, i.e. since $L_{M_w}=L_{\overline{M}}=L_{M_0}$ iff $M_w$ rejects $w$.

Finally \cite{AHU} arrived at a contradiction showing that $L_{M_0}\in\DS(S_2(n))\setminus\DS(S_1(n))$ and thus finishing
the proof.

\vspace{5mm}

After this lengthy but relevant digression we come back to Mr.~Tianrong's paper. In his paper he follows the five step
construction of $M_0$ in \cite{AHU} but with some modifications:
\begin{enumerate}
\item \cite{LBAQ} replaces $S_2(n)$ by $O(S(n))$ for $M_0$ (called $M$ in \cite{LBAQ}) and $S_1(n)$ by $S(n)$ for the $M_x$
(called $M_i$ in \cite{LBAQ}). 
\item \cite{LBAQ} lets $M_0$ be an NTM instead of a DTM.
\item \cite{LBAQ} reorganizes the five steps into six steps as follows:
\begin{enumerate}
\item In \cite{LBAQ} steps 1 and 2 are essentially swapped.
\item The determination of $s$ and $t$ moves from step 3 to step 1. Except for this, negligible changes in notation and
adjustments for differing space bounds, steps 3 and 4 are literally the same word by word.
\item In step 2 of \cite{LBAQ} the space bound $S_2(n)$ of step 1 in \cite{AHU} is replaced by $(1+\lceil\log_2
s\rceil+\lceil\log_2 t\rceil) S(n)$.
\item Step 5 is almost literally the same except for a small introduction: ``By using nondeterminism in $(1+\lceil\log_2
s\rceil+\lceil\log_2 t\rceil) S(n)$ cells''. This is accompanied by a mysterious footnote, explaining that ``$M$ is
somewhat deterministic'', but ``that $M$ we constructed here is a NTM'' (cf. \cite{LBAQ} for the full footnote).
\item Step 1 of \cite{LBAQ} forks of to a step 6 if $x$ is not a code word for a DTM. This step 6 is original. Instead
of simply halting as \cite{AHU} does in this situation, \cite{LBAQ} sets up another simulation and describes at the end,
when this simulation accepts or rejects. Additionally, \cite{LBAQ} starts with: ``Since $x$ is not encoding of some
single-tape DTM.'' After setting up the simulation, \cite{LBAQ} goes on: ``By using its nondeterministic choices, $M$
moves as per the path given by $x$.''
\end{enumerate}
\end{enumerate}
After the construction, \cite{LBAQ} follows the proof by contradiction in \cite{AHU}. Of course, the inequalities need
adjusting and there seems to be an error in calculating the limit in lines 2-4 of p. 8, however this does not seem to
invalidate the overall argument in our point of view. At the end compared to \cite{AHU} a novel argument is
made that the NTM $M$ actually can reverse accepting and rejecting states at the end of its execution: this is indeed
less trivial for NTMs than for DTMs, so this remark is rightfully made, but this issue is resolved by standard results from the
literature.

With the NTM $M$ of \cite{LBAQ} we have $L_M\not\in\DS[S(n)]$ and it is claimed that $M$ uses at most $O(S(n))$ space.
By the standard trick of exponentially enlarging the tape alphabet of $M$ if necessary it is not hard to see,
that $M$ can be constructed to use at most $S(n)$ space. Thus indeed, if everything was correct before, then there is
an $L_d$ with $L_d\in\NS[S(n)]$ but $L_d\not\in\DS[S(n)]$.

This finishes the outline of the core argument for \cite[Thm. 3]{LBAQ} and its blueprint, the proof of \cite[Thm. 11.1.]{AHU}. 

\section{Contradictory consequences of Mr.~Tianrong's approach}
\label{sec:cons}

We assume for the time being in this section that the proof strategy of \cite{LBAQ} was valid. Based on the
modifications of \cite{LBAQ} we make additional modifications. 

We revert modification 2: In our construction $M_0$ will be constructed as a DTM again. Most things should go through
smoothly as the $M_0$ of \cite{AHU} was a DTM anyways, we only have to pay attention to the parts of \cite{LBAQ} where
Mr.~Tianrong invokes nondeterminism. This happens in exactly two places as far as we were able to observe: modification
3(d) and 3(e).

In 3(d) Mr.~Tianrong already admitted that ``$M$ is somewhat deterministic'' (recall: $M$ is the name in \cite{LBAQ} for
$M_0$). It seems that nondeterminism is not essential here. At least Mr.~Tianrong was not able to explain to the author
via emails, why a DTM could not do the same work required to successfully complete step 5. As step 5 in \cite{LBAQ} is
almost word by word identical to step 5 in \cite{AHU} except for the introduction ``By using nondeterminism in
$(1+\lceil\log_2 s\rceil+\lceil\log_2 t\rceil) S(n)$ cells'', 3(d) cannot be considered an obstacle to switching back to
a DTM $M_0$.

In 3(e) something rather strange happens: $x$ is not a valid encoding of a DTM (``Since $x$ is not encoding of some
single-tape DTM.''), but nevertheless this garbage code $x$ should be the program governing the execution of $M_0$
a.k.a.~$M$ later on (``By using its nondeterministic choices, $M$ moves as per the path given by $x$.''). I did not get
an explanation from Mr.~Tianrong on this issue, in particular, he was not able to detail to me how essential
nondeterminism must enter here. It is also completely unclear to me, how step 6 should be useful and what purpose it
serves. Just after explaining the five steps a valid code word $w$ with some additional properties is constructed in
both \cite{AHU} and \cite{LBAQ} and nothing that follows the five steps seems to depend on the behavior of non-code
words. So this will be our second additional modification: we go back to the original construction of \cite{AHU}, which
is simply halting without accepting for any $x$ not a code word.

Overall, this modifications on top of Mr.~Tianrong's modifications remove the nondeterminism and we demonstrated that,
if the argument before our modifications was correct, it is so after our modifications.

With this, we could establish 
\begin{namedthm}{Theorem 3'}
Let $S(n)\geq\log n$ be a space-constructible function. Then there exists a language $L_d$ accepted by a DTM by using
space $O(S(n))$ but by no DTM of space complexity $S(n)$. That is, $L_d\in\DS[S(n)]$ but $L_d\not\in\DS[S(n)]$.
\end{namedthm}

Clearly, the last part is an obvious contradiction. This shows that the assumption that Mr.~Tianrong's argument for
Theorem 3.~was correct must be wrong. Even worse, the whole approach of simply modifying the proof of 
\cite[Thm. 11.1.]{AHU} by replacing faster growth (in form of the pair $S_1(n)$ and $S_2(n)$) by nondeterminism does not work
(at least not without major new ideas, which are absent in the work of Mr.~Tianrong), as it is easily possible to go 
back from NTMs to DTMs as demonstrated above arriving at a fatal contradiction.

\section{The reason for the failure of Mr.~Tianrong's strategy}
\label{sec:cause}

We believe that the source of this contradiction lies in modification 3(c). Mr.~Tianrong changes the space bound for
$M_0$ a.k.a.\ $M$ from $S_2(n)$ to $(1+\lceil\log_2 s\rceil+\lceil\log_2 t\rceil) S(n)$. At this stage the input $x$ is
still allowed to vary freely over all binary words. With changing $x$ the length $n=|x|$ varies and also the associated
DTM $M_x$ (assuming the code word is valid) and with it $s$ and $t$. For what follows let $s(n)$ be the maximum of all
cardinalities of states $s$ for some DTM $M_x$ associated to a code word $x$ of length $n=|x|$ and introduce mutatis
mutandis $t(n)$ for the lengths of encoded tape alphabets. Let $\alpha(n)$ be the analog maximum for the expression
$\lceil\log_2 s\rceil+\lceil\log_2 t\rceil$.

Let's assume $\alpha(n)$ was bounded. Then $s(n)$ and $t(n)$ were both bounded and we could only encode a finite number
of single-tape DTMs; but we need to encode all DTMs which use space no more than $S(n)$ tape cells. However, if we had
$S(n)=0$, corresponding to DFAs, we would already have infinitely many different languages (regular languages) and
therefore infinitely many different automata. The more for $S(n)\geq\log n$. Thus $s(n)$ and $t(n)$ cannot both be
bounded, if we want to prove something about the infinite set of languages $\DS[S(n)]$. But then $(1+\alpha(n))
S(n)\not\in O(S(n))$ or informally ``$(1+\lceil\log_2 s\rceil+\lceil\log_2 t\rceil) S(n)\not\in O(S(n))$'' and this is,
where the error is hidden.

How does \cite{AHU} deal with this issue? They work with $S_2(n)$ instead of $(1+\lceil\log_2 s\rceil+\lceil\log_2
t\rceil) S(n)$, but \cite{AHU} does not require $S_2(n)\in O(S_1(n))$. Rather to the contrary $S_2(n)$ is supposed to
grow faster than $S_1(n)$ in the sense of their infimum condition, thus there is not even an issue for \cite{AHU} to deal
with.

\section{Summary}
\label{sec:sum}

We outlined the proof strategy of \cite{LBAQ} and gave a teleological argument why it cannot work and is beyond
repair. Precisely, if assumed to be correct, slight and logically harmless modifications would yield contradicting
results. We pinpointed the main source of these issues in a simple-minded modification of a correct proof
of \cite[Thm. 11.1.]{AHU} trading faster growth by insinuating to nondeterminism to get a flawed attempt at a proof
of \cite[Thm. 3.]{LBAQ}. Since any new claims made by Mr.~Tianrong all built on his \cite[Thm. 3.]{LBAQ}, all his claims
on proving any new result (Thms. 1., 2. and 3. and Corollaries 1. and 2.) are unjustified and void.
\newpage

Mr.~Tianrong has since updated and reworked his preprint. In particular in the latest version from 4th Aug. 2022 his
main \cite[Thm. 3.]{LBAQ} is now \cite[Thm. 5.]{LBAQv6}. The mysterious footnote containing ``$M$ is somewhat
deterministic'', but ``that $M$ we constructed here is a NTM'' is now gone. Mr.~Tianrong formally addresses the concerns
raised in this note, however in substance he fails to provide any new ideas and to resolve these concerns. His arguments
are essentially unchanged. Therefore all of Mr.~Tianrong original claims in \cite{LBAQv6} are still unjustified and
void.

\section*{Acknowledgment}
The author is grateful for the patience of Mr.~Tianrong in the email discussion in the previous days, although we could
not agree on assessing correctness and quite a lot of things. We also thank Mr.~Tianrong for his invitation to write
this paper and thereby clarifying the mistakes and flaws in Mr.~Tianrong's approach.

\end{document}